\documentstyle[12pt,a4,epsf]{article}
\global\arraycolsep=2pt 
\input{epsf}
\begin{document}
\makeatletter
\def\fmslash{\@ifnextchar[{\fmsl@sh}{\fmsl@sh[0mu]}}
\def\fmsl@sh[#1]#2{%
  \mathchoice
    {\@fmsl@sh\displaystyle{#1}{#2}}%
    {\@fmsl@sh\textstyle{#1}{#2}}%
    {\@fmsl@sh\scriptstyle{#1}{#2}}%
    {\@fmsl@sh\scriptscriptstyle{#1}{#2}}}
\def\@fmsl@sh#1#2#3{\m@th\ooalign{$\hfil#1\mkern#2/\hfil$\crcr$#1#3$}}
\makeatother
\thispagestyle{empty}
\begin{titlepage}

\begin{flushright}
  { \bf  LMU 15/99} \\
  \today
\end{flushright}

\vspace{0.3cm}
\boldmath
\begin{center}
\Large\bf One-Particle Inclusive $B_s \to \bar D_s X$ Decays           
\end{center}
\unboldmath
\vspace{0.8cm}
\begin{center}
  {\large Xavier Calmet} \\
  {\sl Ludwigs-Maximilians-Universit{\"a}t, Sektion Physik,
    Theresienstr. 37, \\ D-80333 M{\"u}nchen, Germany}
\end{center}

\vspace{\fill}

\begin{abstract}
\noindent
We discuss one-particle inclusive $B_s \to \bar D_s X$ decays using a
QCD based method already applied to $B \to \bar D X$. A link between
the right charm non-perturbative form factors of the semi-leptonic
decays and those of the non-leptonic decays is established. Our
results are compatible with current experimental knowledge.
\end{abstract}

\end{titlepage}


\section{Introduction}
Some time ago, a QCD based method was proposed to describe $B\to \bar
D \ell \nu X$ decays, which relies on a short distance expansion (SDE)
and on the heavy quark effective theory (HQET) \cite{BM}.  The
non-perturbative form factors of the singlet operators were
parameterized using the Isgur-Wise function. More recently this method
was extended to one-particle inclusive non-leptonic $B$ decays
\cite{CMS}.  In this case, we have to perform a $1/N_C$ expansion,
which allows to factorize the matrix elements.  One of the goal of
this work is to clarify the link between the matrix elements which
were encountered in the semi-leptonic one-particle inclusive B decays
\cite{BM} and those of the non-leptonic one-particle inclusive B
decays encountered in \cite{CMS}. In fact we shall prove that these
matrix elements are universal. We shall then apply this method to
one-particle inclusive $B_s \to \bar D_s X$ and $B_s \to D_s X$ decays.

It is shown in \cite{CMS} that the one-particle inclusive decays of a
$B$ meson into a vector D meson seem to be, in this framework, well
understood whereas decays of a B meson into a pseudo-scalar D are
troublesome, i.e. the decay widths and spectra for $B \to \bar D^*/D^*
X$ admixtures look to be described correctly, on the other hand the
predictions for $B \to \bar D/ D X$ admixtures decay widths and
spectra do not reproduce the experimental data. Most troublesome is
the fact that the spectra are not even described correctly for large
transfered momentum. According to our method we expect to describe the
experimental data for large transfered momentum particularly well.

Keeping in mind that some problems arose in the description of $B \to
\bar D/D X$ decays, we apply the method developed for these decays to
$B_s \to \bar D_s X$ and $B_s \to D_s X$ decays. The effective
Hamiltonian is identical in both cases.  One-particle inclusive $B_s
\to \bar D_s X$ decay widths have been measured by ALEPH. There are
measurements for semi-leptonic \cite{Aleph1} as well as for
non-leptonic \cite{Aleph2} decays.

The decay rates we are computing can be used to study one-particle
inclusive CP asymmetries in the $B_s$ system \cite{calmet2}, which would
allow an extraction of the weak angle $\gamma$ which is known to be
difficult. This study of $B_s \to D_s X$ decays could also allow to
get a better understanding of the problems encountered in $B \to D X$
decays \cite{CMS}. They are also interesting for experimental physics
especially in the perspective of $B$ factories as the presently available
data on one-particle inclusive $B_s \to D_s X$ decays is sparse.

In the following section, we shall establish the link between the form
factors of the semi-leptonic decays and those of the non-leptonic
decays for the right charm $\bar b \to \bar c$ transition.

\section{From semi-leptonic to non-leptonic decays} 
 We shall consider right charm decays $B \to \bar D X$, i.e. $\bar b
\to \bar c$ transitions. The central quantity in the semi-leptonic
case as well as the non-leptonic case is the function $G$ given by
\begin{equation}
G (M^2) = \sum_X 
\left| \langle B(p_B) | H_{eff} | \bar D(p_{\bar D}) X  \rangle \right|^2
(2 \pi)^4 \delta^4 (p_B - p_{\bar D} - p_X),
\end{equation}
where $| X \rangle$ are momentum eigenstates with momentum $p_X$,
$H_{eff}$ is the relevant part of the weak Hamiltonian and $M^2=(p_B-p_{\bar D})^2$ is
the invariant mass. The states $|X\rangle$ form a complete set,
especially $|X\rangle$ can be the vacuum in the semi-leptonic case,
e.g. $B \to \bar D \ell \nu$ contributes to $B \to \bar D \ell
\nu X$.
This function $G$ is related to the decay rate under 
consideration by 
\begin{equation}
d \Gamma (B \to \bar D X) = \frac{1}{2 m_B} d\Phi_{\bar D}\,\, G(M^2), 
\end{equation}
where $d\Phi_{\bar D}$ is the phase space element of the final state $\bar D$ 
meson.
The relevant weak Hamiltonian is given by
\begin{equation} 
H_{eff} = H_{eff}^{(sl)} + H_{eff}^{(nl)},
\end{equation}
where the semi-leptonic and non-leptonic pieces are given by
\begin{eqnarray}
H_{eff}^{(sl)} &=& \frac{G_F}{\sqrt{2}}  V_{cb} 
(\bar{b} c)_{V-A}( \bar{\ell} \nu)_{V-A} + h.c. ,
\end{eqnarray}
\begin{eqnarray}
H_{eff}^{(nl)} &=&  \frac{G_F}{\sqrt{2}}
 V_{cb} V^*_{ud} \left(  (\bar{b} c)_{V-A}( \bar{u} d)_{V-A}
		+ (\bar{b}T^a c)_{V-A}( \bar{u}T^a d)_{V-A} \right )
           + \\ \nonumber &&
 \frac{G_F}{\sqrt{2}}
 V_{cb} V^*_{cs} \left(  (\bar{b} c)_{V-A}( \bar{c} s)_{V-A}
		+ (\bar{b}T^a c)_{V-A}( \bar{c}T^a s)_{V-A} \right )
+ h.c. ,
\end{eqnarray}
where we have neglected the penguins and the Cabibbo suppressed operators.
The function $G$ can be written as
 \begin{equation}
G (M^2) =\sum_{X} \int d^4 x \,  
        \langle B(p_B)| H_{eff} (x) |\bar D(p_{\bar D}) X \rangle
        \langle \bar D(p_{\bar D}) X |
                H_{eff} (0) | B(p_B)\rangle.
\end{equation}
In the semi-leptonic case we can trivially factorize $G(M^2)$ and obtain
\begin{eqnarray}
  G^{Lep}(M^2) & = &
    \frac{G_F^2}{2} \, |V_{cb}|^2 
      \sum_X (2 \pi)^4 \delta^4 (M - p_X )
      \langle 0    | (\bar \ell \gamma^\mu (1-\gamma_5) \nu)
                   (\bar  \nu \gamma^\nu (1-\gamma_5)  \ell) | 0 \rangle \ \ \ \ \ \ \ \ 
    \\ &&
    \langle B(p_B) | (\bar b \gamma_\mu (1-\gamma_5) c) 
                   |  \bar D (p_{\bar D}) X \rangle
                      \langle \bar D (p_{\bar D}) X
          | (\bar c \gamma_\nu (1-\gamma_5) b) | B(p_B) \rangle.  \nonumber  \\ && \nonumber
\end{eqnarray}
The next steps are to insert heavy quark fields in the effective
Hamiltonian and considering $m_b$ and $m_c$ as large scales, to
perform a SDE as it has been explained in
\cite{BM}. In the leading order of the SDE, $G^{Lep}(M^2)$ reads
\begin{eqnarray}                                     
G^{Lep}(M^2) &=& \frac{G_F^2}{2} |V_{cb}|^2 P_{\mu \nu}^{Lep} (M) \\
&& \sum_{X}  
\langle B(v) |[\bar{b}_{v} \gamma^\mu (1-\gamma_5) c_{v'}] 
|\bar D(v') X \rangle
\langle \bar D(v') X |                   
[\bar{c}_{v'} \gamma^\nu (1-\gamma_5) b_{v}]
                 | B(v)\rangle \,, \nonumber
\end{eqnarray}
where $v$ is the velocity of the $B$ meson, $v'$ the one of the $\bar D$
meson and $P_{\mu \nu}^{Lep}$ is a tensor originating from the
contraction of the lepton fields in the effective Hamiltonian.  This
tensor is given by
\begin{equation}
P_{\mu \nu}^{Lep} (M) = A(M^2) (M^2 g_{\mu \nu} - M_\mu M_\nu) 
                  +B(M^2) M_\mu M_\nu.
\end{equation}
Neglecting the lepton masses, we obtain at tree level
\begin{equation}
A(M^2) = - \frac{1}{3 \pi} \Theta (M^2) \mbox{ and } B(M^2) = 0 \, .
\end{equation}

We shall now consider the non-leptonic case. The non-leptonic case is
more complex because two transitions are possible: the right charm
$\bar b \to \bar c$ transition and the wrong charm one $\bar b \to c
$. The wrong charm transition was treated in \cite{CMS} and we shall
not come back to this issue there since this channel is extremely
suppressed in the semi-leptonic case and was neglected in \cite{BM}
and our aim in this section is strictly to establish the link between
the right charm semi-leptonic and non-leptonic decays. Another
difficulty is that factorization can only be performed in the $1/N_C$
limit. This concept is known to be valuable for non-leptonic
exclusive $B$ mesons decays \cite{Bauer}. In this limit the octet
operators vanish.  Thus we obtain
\begin{eqnarray}
  G^{NL}(M^2) & = &
    \frac{G_F^2}{2} \, |V_{cb} V_{q_1 q_2}^*|^2 \, |C_1|^2 \;
      \sum_X \sum_{X'} (2 \pi)^4 \delta^4 (M - p_X - p_{X'}) \\
    \nonumber &&
    \langle B(p_B) | (\bar b \gamma_\mu (1-\gamma_5) c) 
                   |  \bar D (p_{\bar D}) X \rangle \;
      \langle 0    | (\bar q_1 \gamma^\mu (1-\gamma_5) q_2)
                   | X' \rangle \\
    \nonumber &&
    \langle X' | (\bar  q_2\gamma^\nu (1-\gamma_5) q_1) | 0 \rangle \;
       \langle \bar D (p_{\bar D}) X
          | (\bar c \gamma_\nu (1-\gamma_5) b) | B(p_B) \rangle,
\end{eqnarray}
where the $q_i$'s stand for quarks. We see that assuming that $X$ and
$X'$ are disjoint which is certainly the case in the leading order of
the $1/N_C$ limit, we can at once apply the completeness relation for
$X'$ and we just find our-selves in the same situation as in the
semi-leptonic case.

For the quark transition $b \to c \bar u d$ we have $q_1=u$ and
$q_2=d$, i.e. we have two light quarks whose masses can be neglected
just as the one of the lepton in the semi-leptonic case. We obtain
\begin{equation}
P_{\mu \nu}^{NL} (M) =N_C  P_{\mu \nu}^{Lep} (M),
\end{equation}
where $N_C$ is the color number, and
\begin{eqnarray}                                     
G^{NL}(M^2) &=& \frac{G_F^2}{2} |V_{cb}V_{ud}|^2 P_{\mu \nu}^{NL}(M) \\
&& \sum_{X}  
\langle B(v) |[\bar{b}_{v} \gamma^\mu (1-\gamma_5) c_{v'}] 
|\bar D(v') X \rangle
\langle \bar D(v') X |                   
[\bar{c}_{v'} \gamma^\nu (1-\gamma_5) b_{v}]
                 | B(v)\rangle \,. \nonumber
\end{eqnarray}
The transition $b \to c \bar c s$ can be treated in the same fashion.
In that case the mass of the $c$ quark in the loop cannot be
neglected.  We obtain
\begin{equation}
P_{\mu \nu}^{NL} (M) = A(M^2) (M^2 g_{\mu \nu} - M_\mu M_\nu) 
                  +B(M^2) M_\mu M_\nu,
\end{equation}
where $A(M^2)$ and $B(M^2)$ are given by
\begin{eqnarray} \label{masscquark}
  A(M^2)&=& -\frac{N_C}{3 \pi} \left (1 + \frac{m_c^2}{2 M^2}\right )
 \left (1 - \frac{m_c^2}{M^2}\right )^2 \Theta(M^2-m_c^2),   \\ 
   B(M^2)&=& \frac{N_C}{2 \pi}\frac{m_c^2}{M^2}
   \left(1 - \frac{m_c^2}{M^2}\right )^2 \Theta(M^2-m_c^2), \nonumber 
\end{eqnarray}
at tree level. As explained in \cite{CMS}, we shall set $m_c=1.0\ {\rm GeV}$ to
parameterize the higher order QCD corrections to the current $b \to c
\bar c s$.

We can now establish the connection between the semi-leptonic and the
non-leptonic form factors. The differential decay width for the
semi-leptonic decays is given by
\begin{eqnarray} \label{master}
 \frac{d\Gamma}{dy} &&= \frac{G_F^2}{12 \pi^3}
 | V_{cb} |^2 m_{D}^3   \sqrt{y^2 -1}
\left[(m_{B} - m_{D})^2 E_S (y) \right. \\ 
&& \nonumber \qquad \left.    + (m_{B} + m_{D})^2 E_P (y) 
                    - M^2 \left (E_V (y) + E_A (y)\right) \right], 
\end{eqnarray}
where $y = v \cdot v'$ and where the invariant mass $M^2$ is given by
\begin{eqnarray}
  M^2&=& m_{B}^2 + m_{D}^2 - 2 y m_{B} m_{D}.
\end{eqnarray}  
The differential decay width for the right charm non-leptonic decays
is then given by
\begin{eqnarray} \label{masterNL}
 \frac{d\Gamma}{dy} &=& C_1^2 N_C \frac{G_F^2}{12 \pi^3}
 | V_{cb} V_{ud}^*|^2 m_{D}^3   \sqrt{y^2 -1}
\left[(m_{B} - m_{D})^2 E_S (y) \right. \\ 
&& \nonumber \qquad \left.    + (m_{B} + m_{D})^2 E_P (y) 
                    - M^2 \left(E_V (y) + E_A (y)\right) \right] \\ \nonumber && 
  + C_1^2 \frac{G_F^2}{4 \pi^2}
 | V_{cb}V_{cs}^* |^2 m_{D}^3   \sqrt{y^2 -1}
\left[(B(M^2)-A(M^2))\Big((m_{B} - m_{D})^2 E_S (y) \right. \nonumber \\ 
&& \nonumber \qquad \left.    + (m_{B} + m_{D})^2 E_P (y) \Big) 
                    +A(M^2) M^2 (E_V (y) + E_A (y)) \right], \nonumber 
\end{eqnarray}
where $A(M^2)$ and $B(M^2)$ are given in (\ref{masscquark}).
We see that the right charm semi-leptonic and non-leptonic decay
widths are given in terms of the same form factors
\begin{eqnarray}  \label{etai}
4 m_{B} m_{D} E_S (v \cdot v') &=& 
\sum_{X}      \langle B(v) | [\bar{b}_{v} c_{v'}] 
|\bar D(v') X \rangle
\langle \bar D(v') X |
[\bar{c}_{v'} b_{v}]
                               | B(v) \rangle   \\
- 4 m_{B} m_{D} E_P (v \cdot v') &=& \sum_{X} 
      \langle B(v) | [\bar{b}_{v} \gamma_5 c_{v'}]
|\bar D(v') X \rangle
\langle \bar D(v') X |                           
[\bar{c}_{v'} \gamma_5 b_{v}]
                         | B(v) \rangle   \nonumber \\
4 m_{B} m_{D} E_V (v \cdot v') &=& \sum_{X} 
      \langle B(v) | [\bar{b}_{v} \gamma^\mu c_{v'}] 
|\bar D(v') X \rangle
\langle \bar D(v') X |                           
[\bar{c}_{v'} \gamma_\mu b_{v}]
                         | B(v)\rangle   \nonumber \\
4 m_{B} m_{D} E_A (v \cdot v') &=& \sum_{X} 
      \langle B(v) | [\bar{b}_{v} \gamma^\mu \gamma_5 c_{v'}] 
|\bar D(v') X \rangle
\langle \bar D(v') X |                           
[\bar{c}_{v'} \gamma_\mu \gamma_5 b_{v}]
                         | B(v)\rangle  . \nonumber 
\end{eqnarray}
One important point should be stressed. This set (\ref{etai}) of
non-perturbative form factors describes a transition from
a $B$ meson into a state with a $D$ meson whatever the intermediate
state might be.  It has been shown in \cite{BM} that we can determine
these matrix elements in the semi-leptonic case using constraints from
the heavy quark symmetry (HQS) and a saturation assumption. These
non-perturbative form factors were given in \cite{BM} for each single
decay channel. So the non-leptonic right charm $B \to \bar D X$ decays
can be deduced from the semi-leptonic ones. Note that we have
neglected the renormalization group improvement which had been
considered in \cite{BM} since this effect is small. Therefore we set $C_{11} =
C_3=1$ and $C_{18}=0$ in the set of non-perturbative form factors
given in \cite{BM}.

After the connection between the non-leptonic and the semi-leptonic
case has been established, we shall consider $B_s \to \bar D_s X$ and
$B_s \to D_s X$ decays.
\section{The decays $B_s \to \bar D_s X$ and $B_s \to D_s X$}
As mentioned previously the effective weak Hamiltonian is identical to
the one of the $B \to \bar D X$ case, therefore the equations
(\ref{master}) and (\ref{masterNL}) do also describe the right charm
decay of a $B_s$ meson into a $\bar D_s$ meson if one replaces $m_B$
by $m_{B_s}$ and $m_D$ by $m_{D_s}$.  We have a new set a
non-perturbative form factors
\begin{eqnarray}  \label{etainew}
4 m_{B_s} m_{D_s} E_S (v \cdot v') &=& 
\sum_{X}      \langle B_s(v) | [\bar{b}_{v} c_{v'}] 
|\bar D_s(v') X \rangle
\langle \bar D_s(v') X |
[\bar{c}_{v'} b_{v}]
                               | B_s(v) \rangle   \\
- 4 m_{B_s} m_{D_s} E_P (v \cdot v') &=& \sum_{X} 
      \langle B_s(v) | [\bar{b}_{v} \gamma_5 c_{v'}]
|\bar D_s(v') X \rangle
\langle \bar D_s(v') X |                           
[\bar{c}_{v'} \gamma_5 b_{v}]
                         | B_s(v) \rangle   \nonumber \\
4 m_{B_s} m_{D_s} E_V (v \cdot v') &=& \sum_{X} 
      \langle B_s(v) | [\bar{b}_{v} \gamma^\mu c_{v'}] 
|\bar D_s(v') X \rangle
\langle \bar D_s(v') X |                           
[\bar{c}_{v'} \gamma_\mu b_{v}]
                         | B_s(v)\rangle   \nonumber \\
4 m_{B_s} m_{D_s} E_A (v \cdot v') &=& \sum_{X} 
      \langle B_s(v) | [\bar{b}_{v} \gamma^\mu \gamma_5 c_{v'}] 
|\bar D_s(v') X \rangle
\langle \bar D_s(v') X |                           
[\bar{c}_{v'} \gamma_\mu \gamma_5 b_{v}]
                         | B_s(v)\rangle  . \nonumber
\end{eqnarray}
Once again we can find a parameterization for these non-perturbative
form factors using the semi-leptonic decays. We shall consider the $s$
quark as being massless and we can therefore use the very same heavy
quark symmetry relations as in the case $B \to \bar D X$. As it has
been argued in \cite{BM}, the HQS implies that at $v \cdot v' =1$ the
inclusive rate is saturated by the exclusive decays into the lowest
lying spin symmetry doublet $\bar D_s$ and $\bar D_s^*$.  The $\bar
D_s^*$ subsequently decays into $\bar D_s$ mesons and thus at $v \cdot
v' = 1$ the sum of the exclusive rates for $B_s \to \bar D_s \ell^+
\nu$ and $B_s \to \bar D_s^* \ell^+ \nu$ is equal to the one-particle
inclusive semi--leptonic rate $B_s \to \bar D_s \ell^+ \nu X$.  Making
use of this assumption and of the spin projection matrices for the
heavy $B_s$ and $\bar D^{(*)}_s$ mesons, we obtain:
\begin{eqnarray}
E_{i}(v \cdot v') &=& \frac{1}{16}
|\mathrm{Tr}
\{\gamma_{5}(1 + \fmslash{v}) \Gamma_{i}(1 + \fmslash{v'})\gamma_{5} \}|^{2}
|\xi(y)|^2 \\ &&
+
\frac{1}{16} \sum_{Pol}
|\mathrm{Tr}
\{\gamma_{5}(1 + \fmslash{v}) \Gamma_{i}(1 + \fmslash{v'})\fmslash{\epsilon}
\}|^{2} |\xi(y)|^2 \mathrm{Br}(\bar D_s^* \to \bar D_s X), \nonumber 
\end{eqnarray}
where $i$ stands for $S,P,V$ or $A$, the sum is over the polarization
of the $D^*$ meson and $\xi(y)=1-0.84(y-1)$ is the Isgur-Wise function
measured by CLEO \cite{CLEO}. The branching ratio $\mathrm{Br}(\bar
D_s^* \to \bar D_s X)$ is the new input and since a $ D_s^{*-}$
always decays into a $ D_s^-$, we have $\mathrm{Br}(\bar D_s^* \to
\bar D_s X)=100 \%$.  We then obtain
\begin{eqnarray}\label{B0D-}
E^{B_s^0D_s^-}_S (y) &=& \frac{1}{4}(y+1)^2 |\xi(y)|^2   
                     \\ \nonumber 
E^{B_s^0 D_s^-}_P (y) &=& \frac{1}{4} (y^2 - 1) |\xi(y)|^2 
              \\ \nonumber 
E^{B_s^0 D_s^-}_V (y) &=&  \frac{1}{2}(y+1)(2-y) |\xi(y)|^2 
\\  \nonumber 
E^{B_s^0 D_s^-}_A (y) &=& - \frac{1}{2} (y+2) (y+1) |\xi(y)|^2.
\end{eqnarray}
The non-leptonic decays $B_s \to \bar D_s X$ can be calculated using
these non-perturbative form factors. It is clear that this saturation
assumption is a crude approximation, but it is well motivated by the
HQS at $y=1$ and the available phase space is not very large, this has to be
treated as a theoretical uncertainty due to non-perturbative physics.
The results obtained for the semi-leptonic decays rates in $B \to \bar
D X \ell \nu$ \cite{BM} give us some confidence in our method.

We shall now consider the wrong charm decays of a $B_s$ meson. They
are induced by the quark transition $\bar b \to c$.  The wrong charm
$B^0_s \to D^{*+}_s X$ decay width can be estimated using the method described in
\cite{CMS}, which corresponds to a rescaling of the parton
calculation. In the leading order of the $1/N_C$ and of the
$1/m_{B_s}$ expansions, the differential decay width reads
\begin{eqnarray}
  \frac{d\Gamma}{dy} &=& \frac{3 G_F^2 C_1^2}{2 \pi^3 M^2}  \sqrt{y^2-1} \
  m_{D_s}^3 |V_{cb} V_{cs}^*|^2  y \left(M^2-m_{D_s}^2\right)^2
  \Theta(M^2-m_c^2) F,
 \end{eqnarray}
 where $F$ is a channel dependent non-perturbative form factor. We have
 \begin{eqnarray}
   F^{B_s^0 D_s^+}=f \left(1+3 \Gamma(D_s^* \to D_s X') \right)= 4 f,
   \end{eqnarray}
   where $X'$ is a pion or a photon and $f$ is the constant defined in
   \cite{CMS}; we had $f=0.121$.  Note that the wrong charm decay is
   being modeled and we have restricted our-selves to the so-called
   model 2 of \cite{CMS} since this model seems to yield better
   results than model 1.

\section{Discussion of the results}

In table $1$, we compare our predictions with the experimental data
found in \cite{PDG}. In the semi-leptonic case the method yields
results which agree with the data. Note that we have considered the
$\tau$ lepton as being massive.  On the other hand, it is not clear if
the non-leptonic decays are problematic, our results are in the
experimental error range though at the inferior limit. One should keep
in mind that we had estimated in \cite{CMS} that corrections to our
calculation could be fairly large and in the worst case up to $30\%$.
It would be interesting to measure the rate $\Gamma(B_s \to \bar
D^{*-}_s X)$ to test the agreement between theory and experiment in
this channel. Remember that for the decays $B \to \bar D/D X$
described in \cite{CMS}, theory and experiment looked to be in
agreement for the $B \to \bar D^*/D^* X$ decays and in disagreement
for $B \to \bar D/D X$ decays although this could be accidental, for a
discussion of this problem see \cite{CMS}.

Data is sparse on one-particle inclusive $B_s$ decays, especially no
spectra are available. It would be instructive to compare the spectra
to check if the same discrepancy appears as in \cite{CMS}, where the
spectra for the $B \to \bar D^*/D^* X$ meson decays seemed to be
described correctly and on the other hand the spectra for the decays
of a $B \to \bar D/D X$ were not compatible with the experimental
data, especially at the non recoil point where the method should work
at its best, this effect being therefore very difficult to understand.
Although the extension of the method developed for one-particle
inclusive $B$ decays to $B_s$ decays is trivial, the results we have
obtained are interesting especially in the perspective of $B$
factories. These results could also be used to study mixing induced
one-particle inclusive CP asymmetries in the $B_s$ system
\cite{calmet2}, this allows to determine the weak angle $\gamma$,
which is known to be very difficult.

If the problems encountered in the one-particle inclusive $B$ decays
\cite{CMS} were not present in $B_s$ decays, one could constrain the
kind of diagrammatic topologies contributing to the one-particle
inclusive $B$ decays. In $B$ decays as well as in $B_s$ decays we have
assumed that the dominant diagrammatical topology contributing to the
right charm decay rates is spectator like.  This study of $B_s$ decays
once confronted to more precise experimental results could allow to
test the influence of the light spectator quark.
\begin{table}  \label{T1} 
\begin{center}
\begin{tabular}{|l|l|l|} 
\hline
Mode & Br (theory) & Br (data from \cite{PDG}) \\
\hline
$B_s^0 \to D_s^- X$                   & $64.9 \%$ & $  (92 \pm 33)   \%$ \\
$B_s^0 \to D_s^+ X$                   & $3.3 \%$ & $ $ \\
$B_s^0 \to D_s^- \ell^+ \nu X$              & $9.1\%$  & $ (8.1 \pm 2.5)\%$ \\
$B_s^0 \to D_s^{-} \tau^+ \nu_{\tau} X$               & $2.7 \%$ & $  $ \\
$B_s^0 \to D_s^{* -}X$                       & $49.6\%$  & $   $ \\
$B_s^0 \to D_s^{* +}X$                       & $2.5\%$  & $   $ \\
$B_s^0 \to D_s^{* -} \ell^+ \nu X$              & $7\%$  & $   $ \\
$B_s^0 \to D_s^{* -} \tau^+ \nu_{\tau} X$               & $2 \%$ & $  $ \\ 
\hline
\end{tabular}  
\end{center}
\caption{Comparison of our results with data. To get branching ratios,
         we used $\tau_{B^0_s} = 1.55 $ ps.}
\end{table} 
\section{Conclusions}
We have clarified the link between the non-perturbative form factors
of the semi-leptonic and non-leptonic $B \to \bar D X$.  We have
applied a method described in \cite{BM} and \cite{CMS} to
semi-leptonic and non-leptonic $B_s \to \bar D_s X$ and $B_s \to D_s
X$ decays, this can be done easily by modifying the saturation
assumption. It is too early to see if the same problems which were
encountered in \cite{CMS} do also appear in our case, the reason being
the lack of experimental data. Our results are compatible with current
experimental knowledge.

\section*{Acknowledgment}
The author is grateful to Professor L. Stodolsky for his hospitality
at the ``Max-Planck-Institut f\"ur Physik'' where this work was
performed. He would like to thank Dr. Z.Z. Xing for reading this
manuscript and for his encouragements to publish the present results
and Dr. A. Leike for is very useful comments.

\end{document}